\documentclass[preprint2]{aastex631}  

\usepackage{pgfplots}

\begin{document}

\def\bprp{\mbox{$(G_{BP}-G_{RP})$}}
\def\gg{\mbox{$G$}}
\def\bp{\mbox{$G_{BP}$}}
\def\rp{\mbox{$G_{BP}$}}

\newcommand{\vdag}{(v)^\dagger}
\newcommand\aastex{AAS\TeX}
\newcommand\latex{La\TeX}

\graphicspath{{./}{figures/}}

\shortauthors{Denilso Camargo}

\title{Multiple machine-learning as a powerful tool for the star clusters analysis}

\correspondingauthor{Denilso Camargo}
\email{denilso.camargo@gmail.com}

\author{Denilso Camargo}
\affil{Col\'egio Militar de Porto Alegre, Minist\'{e}rio da Defesa - Ex\'{e}rcito Brasileiro\footnote{Civil staff} \\
Av. Jos\'{e} Bonif\'{a}cio 363, Porto Alegre, 90040-130, RS, Brazil}

\begin{abstract}
This work proposes a multiple machine learning method (MMLM) aiming to improve the accuracy and robustness in the analysis of star clusters. The MMLM performance is evaluated by applying it to the reanalysis of the old binary cluster candidate - NGC 1605a and NGC 1605b - found by \citet{Camargo21} (hereafter C21).
The binary cluster candidate is analyzed by employing a set of well established machine learning algorithms applied to the Gaia-EDR3 data. Membership probabilities and open clusters (OCs) parameters are determined by using the clustering algorithms pyUPMASK, ASteCA, Kmeans, GMM, and HDBSCAN. In addition, a KNN smoothing algorithm is implemented to enhances the visualization of features like overdensities in the 5D space and intrinsic stellar sequences on the color-magnitude diagrams (CMDs).
The method validates the clusters' parameters previously derived, however, suggests that their probable members-stars are distributed over a
wider overlapping area. 
Finally, a combination of the elbow method, t-SNE, kmeans, and GMM algorithms group the normalized data into 6 clusters, following C21.
In short, these results confirm NGC1605a and NGC1605b as genuine OCs and reinforce the previous suggestion that they form an old binary cluster in an advanced stage of merging after a tidal capture during a close encounter. Thus, MMLM has proven to be a powerful tool that helps to obtain more accurate and reliable clusters parameters and its application in future studies may contribute to a better characterization of the Galaxy's star cluster system.

\end{abstract}


\section{Introduction} \label{sec:intro}

The current era is characterized by the increasing use of artificial intelligence (AI) that is radically transforming everyday life, particularly the way humans acquire and process information. An AI system is formed by the combination of several machine learning algorithms, not just a single one.   Machine learning (ML) is a section of AI that allows the system to autonomously learn from data making them more effective of solving complex problems. 
ML techniques have been increasingly used in astronomy research, mainly those related to star clusters. Such approaches are implemented to provide valuable memberships, accurate parameters, and to search for unknow clusters, especially open clusters \citep[][and references therein]{Li25}.   

Open clusters are key laboratories for improving our knowledge on the Galaxy's structure, dynamics, chemical properties, formation and evolution, since it seems that they are the preferred mode of star formation \citep{Camargo11, Majaess13, Hao21, Hou21, Dobbs22, Rieder22}. In the early evolutionary stage OCs are known as embedded clusters (ECs) and are often found composing star  cluster aggregates \citep{Camargo11, Camargo12, Camargo15a, Camargo15b, Kuhn19, Karam22, Badawy24, Guimei25}. In this sense, \citet{Camargo16} suggest that giant molecular clouds may collapse into several
clumps forming multiple systems that they named embedded cluster aggregates. 

However, although multiple systems are common in the EC stage \citep{Kovaleva20}, in the evolved phase of OC these multiplets are rare \citep{Fuente09, Conrad17,Liu19, Angelo21, Li24, Ishchenko24}. The scarcity of such aggregates at a late evolutionary stage suggests that they are short-lived and most of their member-ECs merge early or end up completely disrupted.

\begin{figure}
\centering
\begin{minipage}[b]{1.0\linewidth}
\begin{minipage}[b]{0.48\linewidth}
\resizebox{\hsize}{!}{\includegraphics{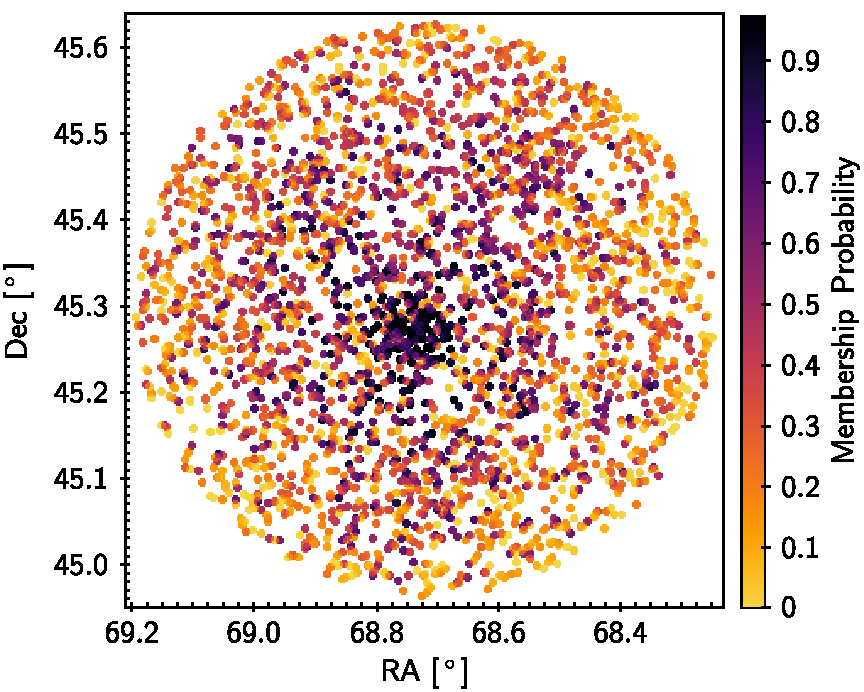}}
\end{minipage}\hfill
\begin{minipage}[b]{0.498\linewidth}
\includegraphics[width=\textwidth]{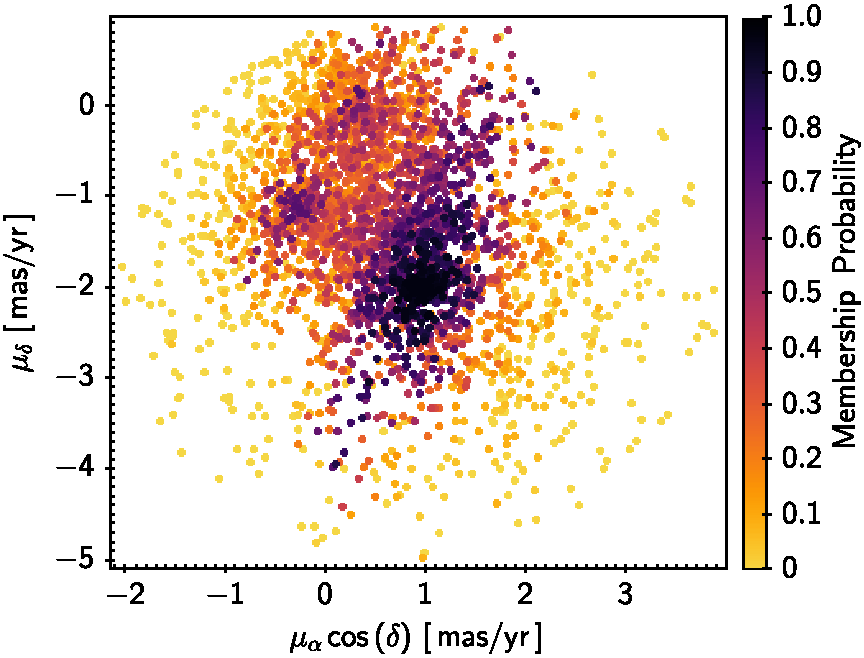}
\end{minipage}\hfill
\caption{Spatial and PM distributions of stars for an extraction area of $R=20'$ centered in the coordinates of NGC1605a and color coded by the pyUPMASK membership probabilities. Both diagrams reveal multiple substructures, especially the PM distibution.}
\label{z1b1}
\end{minipage}
\end{figure}

Alternatively, cluster pairs may be formed occasionally by tidal capture of clusters with different ages \citep{Fuente09}. In this sense, C21 reported the discovery of the first old OC pair within the Galaxy. In fact, such a binary open cluster is composed by an old and an intermediate-age object. The discovery was initially based on the 2MASS near-infrared photometry and subsequently confirmed by using the visual photometry and astrometry provided by Gaia EDR3. An OC (FSR 694) has been found by \citet{Froebrich07} in this region, however, for simplicity purpose the binary cluster's components were named NGC1605a and NGC1605b.

\begin{figure*}
\centering

\resizebox{\hsize}{!}{\includegraphics[width=0.3\textwidth]{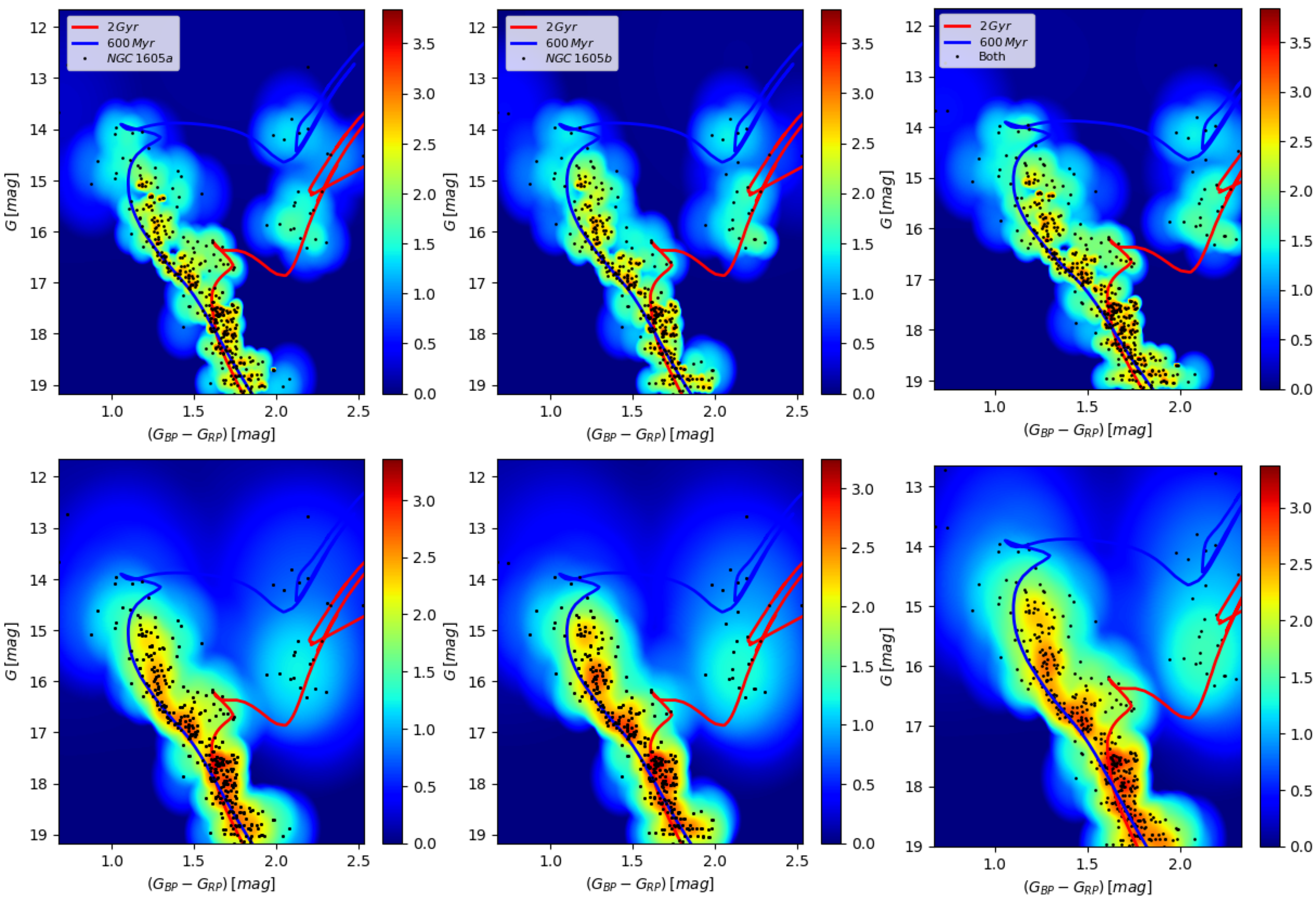}}

\caption{Gaia-EDR3 optical CMDs for the probable member-stars (black circles) within an extraction area of $R=60'$ centered in the coordinates of NGC1605a (\textit{left panels}), NGC1605b (\textit{middle panels}), and a combination of both (\textit{right panels}).  The heatmaps are built by applying the KNN-smoothing algorithm smoothing on $K=3$ (\textit{top panels}) and $K=8$  neareast neighbors (\textit{bottom panels}) and are color coded by the smoothness level. The two clusters are present in both CMDs and are highlighted by the isochrones fitting and the smoothing graphs.}
\label{z1c}

\end{figure*}

\begin{figure*}[!t]
\centering
\begin{minipage}[b]{0.70\linewidth}

\resizebox{\hsize}{!}{\includegraphics{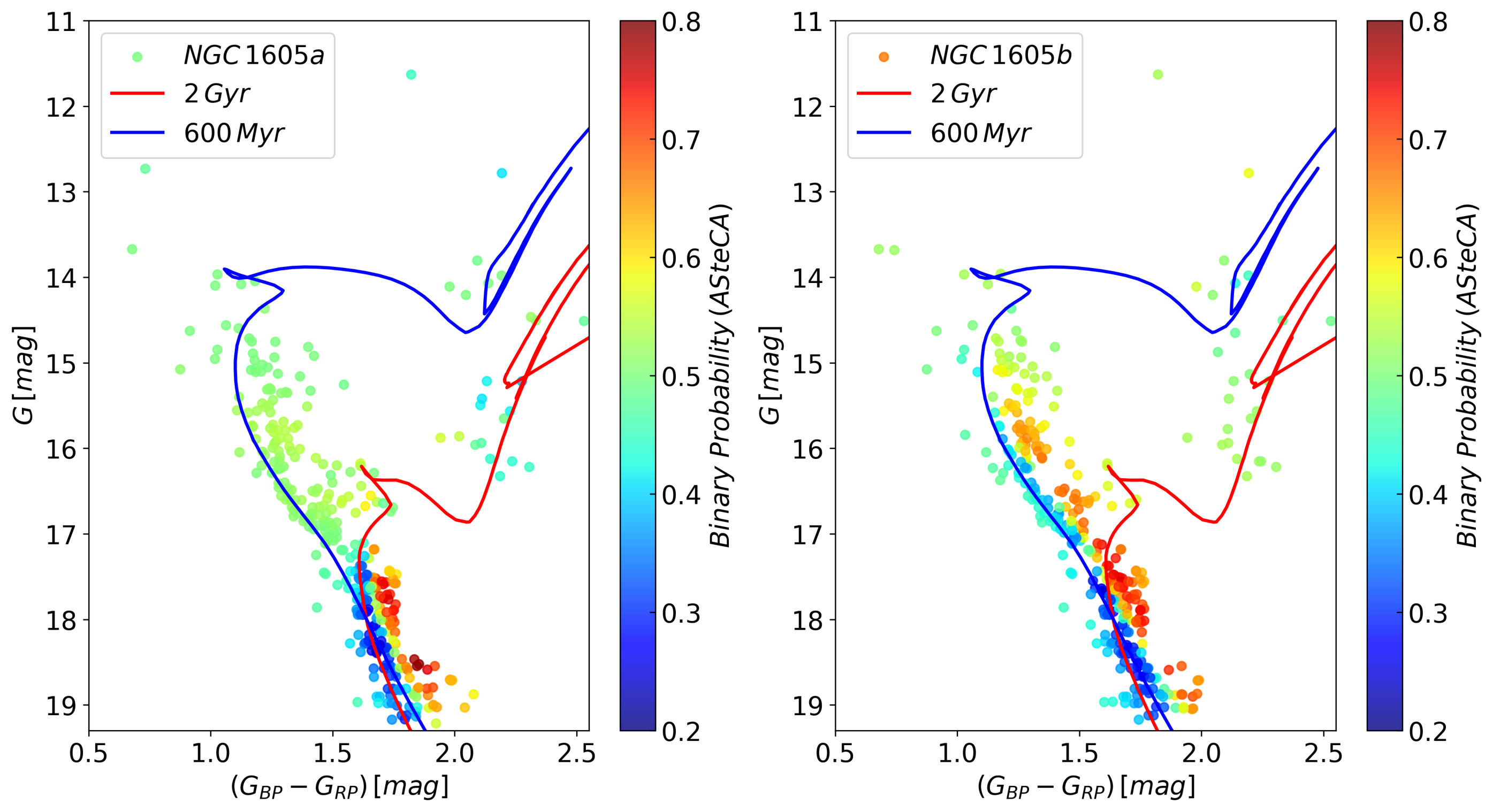}}

\end{minipage}\hfill

\caption{Gaia-EDR3 CMDs for the NGC1605a (\textit{left panel}) and NGC1605b memberlists  (\textit{right panel}). The color bar denotes the binary probability for the cluster members candidates. The bulk of probable binary stars in both CMDs are located in the region corresponding to the NGC1605a upper MS.}

\label{z1b}
\end{figure*}

Thanks to the advent of sky surveys \citep{Skrutskie06, Minniti10, Wright10, Saito24} several new star clusters were found \citep[e.g.,][]{Froebrich07, Borissova11, Camargo16, Camargo18, Camargo19, Garro20, Garro22, Minniti21a, Minniti21b}. Furthermore, the Gaia source list combined with the application of some clustering algorithms have allowed the identification of numerous previously undetected OCs \citep{Cantat18,  Liu19, Castro20, Li22, Li23, Castro22, He22a, He22b, Hunt23, Chi23, Qin23}. Amidst these discoveries, some binary clusters were also identified \citep[see,][]{Conrad17, Bisht21, Casado21, Song22, Chen24, Li24, cinar25, Palma25, Hu25, Qin25}. For illustration purposes, by using Gaia EDR3 \citet{Angelo21} point out that Ruprecht 100 and Ruprecht 101 form an intermediate age binary cluster, \citet{Piatti22} suggest an on-going collision between IC 4665 and Collinder 350, and \citet{Li24}  provide a catalog with 13 new findings.

Nonetheless, its enormous legacy, as well as previous surveys, Gaia also presents photometric and astrometric limitations. There is a significant effort aiming at improvements, however, in order to  avoid noises in clusters' derived parameters and/or to uncover their true natures, some care is recommended when using Gaia EDR3 catalogue \citep{Gaia21, Fabricius21, Riello21}. For instance, the parallax zero point offset may produce a large systematic error in the distance determination, especially for distant sources \citep{Lindegren21, Huang21, Li22, Riess22, Khan23, Molinaro23}.

Thus, aiming to garantee the accuracy of both - clusters' membership lists and the derived parameters - it is relatively common in the literature to adopt some quality cuts in the Gaia photometric and astrometric data \citep{Cantat18, Castro22, Penoyre22a}. The main quality cuts commonly applied to the Gaia sources are the following: (\textit{i}) a cut based on the ‘Renormalized Unit Weight Error’, $RUWE < 1.4$, that remove most of the unresolved binaries and poor quality astrometric solutions, and (\textit{ii}) so as to avoid large uncertainties, a photometric cut is used excluding stars fainter than $G=18\,mag$. These cuts ensure that the most of the detected overdensities are true clusters instead of field flutuations.

On the other hand, these filters sacrifice the completeness of some cluster candidates by removing true members, such as faint sources and binary stars. Thus, these quality cuts may impact the detectability of clusters and its membership lists, especially for very young and  distant and/or faint old OCs. \citet{Buckner24} argue that, for distant OCs, these cuts could artificially remove from the data cluster's substructures and/or true subclusters. As a result some true OCs could be being missed or merged together into a single cluster. They also verified that the cut for faint magnitudes, applied to the clusters identified in \citet{Cantat18}, removed significantly the number of true members, while provide only a marginal effect in the non-members.
In this sense, \citet{Perren22} considered only the proper motion to assign the membership probability for very distant clusters. They argue that for such clusters these extra data tend to add more noise than information, especially the parallax \citep{Chi23}. It is expected that the main sequence (MS) stars for distant and/or old OCs are mostly fainter than 18 mag that is the quality cut commonly adopeted for Gaia \citep{Chen24}. In line with this, \citet{Piatti22} suggested a careful reanalysis of the undetected old OCs in the MWSC catalogue \citep{Anders21}, due to the magnitude cut used for faint stars.

\begin{figure}
\centering
\begin{minipage}[b]{1.0\linewidth}

\includegraphics[width=\textwidth]{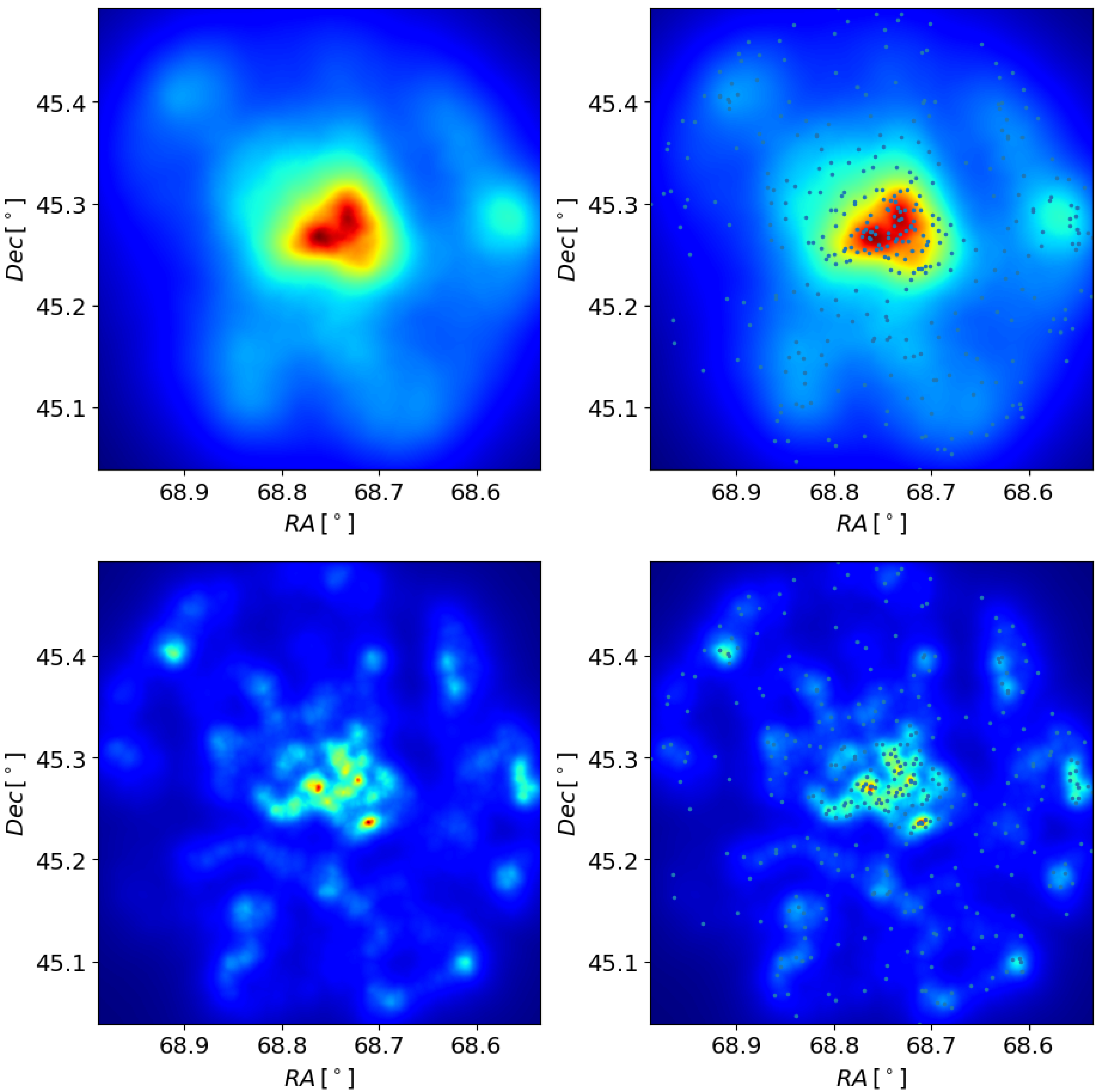}

\caption{Heatmaps of the spatial distribution for stars in the NGC1605b memberlist, built by using the KNN-smoothing algorith smoothing on $K=28$ (\textit{top panels}) and $K=5$ (\textit{bottom panels}) neareast neighbors. The \textit{right panels} display the schematic distribution of the cluster probable member-stars as black circles overlaid on the heatmaps.}
\label{g}
\end{minipage}\hfill
\end{figure}

In addition, most OCs are found in the Galactic disk close to the midplane and, thus, crowding and extinction may affect the sources increasing both astrometric and photometric errors. As an optical survey the Gaia data is vulnerable to extinction \citep{Babusiaux23, Buckner24}.

The paper is structured as follows. 
In Sect.~\ref{sec:2a} the Gaia-EDR3 dataset and the methods used in this study are described.
Sect.~\ref{sec:3} displays the results and discussion. Finally, Sect.~\ref{sec:4} shows the concluding remarks.

\section{Methods}
\label{sec:2a}

This study provides a method dedicated to star cluster analysis - the MMLM - and tests its effectiveness through the reanalysis of a binary cluster candidate, NGC1605a and NGC1605b. The analysis is based on the Gaia-EDR3 data \citep{Gaia23} by employing a set of both supervised and unsupervised machine learning algorithms, which are useful tools to uncover hidden patterns and groupings within data. 
Basically, machine learning algorithms investigate the data's structure by grouping its points into distinct clusters based on similarities. 
These algorithms can help reveal patterns and structures like star cluster in the Gaia data, substructures within a star cluster, and multiple clusters \citep{Qin23, Chen24, Li24, Wu24}.

There are several methods available for the OCs' exploration and, each approach has its own advantages and limitations depending on the research nature and objectives. Thus, given the complexity of some star clusters' analyses, it was decided to implement a combination of a set of machine learning, as the use of different types of tools may  add value to results. Machine learning combination is relatively common in astronomical research \citep{Agarwal21, Noormohammadi23, Qin23, Li24b}. 

The old binary cluster is reanalized by using the unsupervised clustering algorithms HDBSCAN (\textit{Hierarchical Density-Based Spatial Clustering of Applications with Noise}), GMM (\textit{Gaussian Mixture Model}), and Kmeans\footnote{Classical and Fuzzy implementations.}. The ASteCA (\textit{Automated Stellar Cluster Analysis}) and pyUPMASK (\textit{python implementatin of the unsupervised photometric membership assignment in stellar clusters}) also are used. This study also makes use of the supervised machine learning algorithms KNN (\textit{K-nearest Neighborhood})  and t-SNE (\textit{t-Distributed Stochastic Neighbor Embedding}).

The pyUPMASK algorithm \citep{Krone14, Pera21} is an unsupervised clustering method for star clusters that provides membership probabilities.

The ASteCA \citep{Perren15, Perren20} derives the cluster fundamental parameters by comparison of the clusters intrinsic sequences with a synthetic cluster CMD built based in such sequences. The algorithm find the best fitting PARSEC isochrone \citep{Bressan12} to the outliers-free CMD.

GMM \citep{Dempster77} is a probabilistic clustering algorithm, which considers that the data points are distributed as a finite number of Gaussians, with each one of them corresponding to a cluster. It is a powerful tool for providing the cluster membership probability \citep{Hunt23}.

The HDBSCAN \citep{Campello13, McInnes17} build an hierarchical condensed tree by repeatedly merging close together points. Overdensities in the condensed tree are clusters. Such an hierarchical procedure is a powerful tool to isolate outliers. HDBSCAN can identify clusters of varying density.

Kmeans split the data in clusters by find k centroids, determined a priori. Basically, data points (i.e., stars) are clustered around a k number of cluster centers by using the distance (usually Euclidean) between each one to the centroids. Points are clustered to the nearest centroid.

KNN (k nearest neighbors) is a supervised machine learning that makes predictions based on the distance between nearest neighbors \citep{Cover67}. The tool assumes that similar points are close to each other in the feature space. KNN approaches are commonly used combined together with another algorithm.  
A KNN smoothing algorithm is implemented with the aim of building heatmaps to uncover stellar sequences and identify areas with high concentration. Such a tool is able to identify trends and patterns in data, even when some values are missing.

\begin{figure}
\centering
\begin{minipage}[b]{0.7\linewidth}
\includegraphics[width=\textwidth]{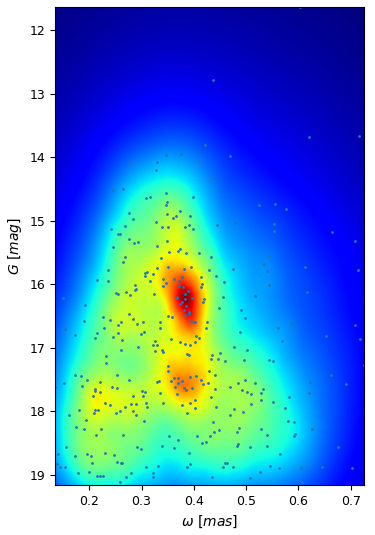}
\end{minipage}\hfill
\caption{ Parallax versus G magnitude for the NGC1605b memberlist. The heatmaps are built by the KNN-smoothing algorithm.}
\label{z1y}

\end{figure}

The t-SNE is another useful technique \citep{Hinton03, van08} that, in this study, is employed for the visualization of the Gaia 5D astrometric space by reducing it to a 2D space. This approach has a main parameter that governs the data distribution, the perplexity.

Following C21, Gaia data was extracted for two OCs candidates, NGC1605a and NGC1605b.
This previous work found, in the NGC1605 region, two stellar populations with discordant ages and similar kinematics, arguing for a close encounter of two clusters, which are in the final stage of an ongoing merging.

As a first step, the set of tools is applied to the raw Gaia data (R=20') to estimate the PM components for both clusters, without applying any quality cut. Then, new extractions are carried out by using as quality cut a proper motion radius of $R=1\,\,mas\,yr^{-1}$, centered in the derived PM central coordinates ($\mu_{\alpha}\,cos(\delta)$, $\mu_{\delta}$) of each cluster.
Additionally, the errors for both proper motion coordinates are limited to less than $0.4\,\,mas\,yr^{-1}$ and $RUWE<1.4$.
It is important to point out that the adopted PM radius cut is shorter than the common value for most single Galactic OCs. Although this procedure certainly affects the clusters' completeness, it is necessary to demonstrate that there are at least two OCs experiencing an merger event, as suggests C21.   
The PM quality cut adopted provides an initialy outliers decontamination, however, it may discard clusters' members ejected from them as an effect of the close encounter, since these stars differ significantly from the kinematics of the bulk of clusters' members.
The purpose of this restriction is to build a dataset that approximates what will be the merger final product, while still preserving some characteristics of the clusters as individual entities. 
As the ASteCA points to extended structures, extraction radii of $R=60'$ are adopted.
Subsequently, the extracted Gaia dataset is submitted to the set of clustering algorithms (pyUPMASK, ASteCA, GMM, and HDBSCAN) individually. The stellar content assigned with membership probability higher than $50\%$ in all algorithms are considered as probable cluster members and compose the final dataset.

In addition to the quality cuts aformentioned, large parallax and PMs errors are another important limitation for identifying faint clusters' member-stars. Then, the errors are not used in the membership determination.

\begin{figure}
\centering
\begin{minipage}[b]{0.7\linewidth}
\includegraphics[width=1.0\textwidth]{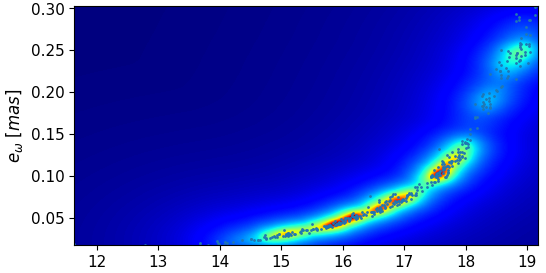}
\includegraphics[width=1.0\textwidth]{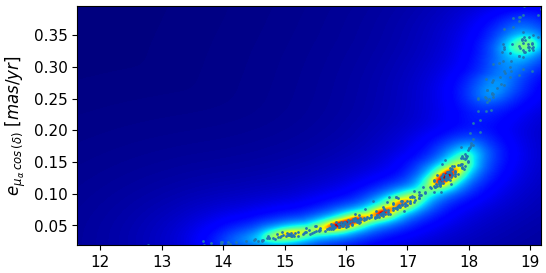}
\includegraphics[width=1.0\textwidth]{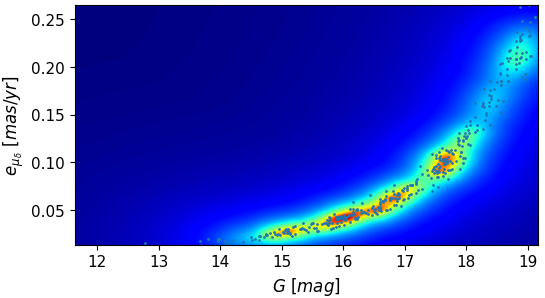}
\end{minipage}
\caption{Parallax and PMs errors versus G magnitude for stars in the NGC1605b memberlist.   
}
\label{f16}
\end{figure}

\subsection{Astrophysical parameters}
\label{sec:2b}

The clusters' ages and heliocentric distances are derived by ﬁtting PARSEC isochrones \citep{Bressan12} on the CMDs built with the list of probable members of each cluster. The heliocentric distance is estimated by using $A_G\,=\,2\times{E(G_{BP}-G_{RP})}$ and $R_V=3.1$.
In addition, the heliocentric distances are computed as the inverse of the Gaia EDR3 parallax, after applying the zero-point corrections \citep{Lindegren21}.

The structural parameters for the present objects are uncovered by fitting a King's profile \citep{King62} on the radial density profile (RDP), which are built with stars from the probable memberlist of each object.
The parameters derived by fitting a King's law (eq.~\ref{a}) are the central density of stars (${\sigma}_0$), the core radius ($R_{c}$), and the tidal radius ($R_{t}$). The parameter ${\sigma}_0$ represents the projected density of stars at the cluster center. The $R_{c}$ defines the radius at which the projected density of cluster member-stars reaches half the central ones, while the $R_{t}$ denotes the radius at which the density drops zero.

The model was initially proposed to describe the globular clusters structure, but later it also became an important tool to derive structural parameters for OCs with a reliable accuracy \citep[e.g][and references therein]{Camargo10, Camargo11, Camargo12, Camargo13, Camargo16}. In the case of OCs more attention is required, since external perturbations - such as Galactic tidal stress, another clusters and/or molecular clouds flybys - may lead the OCs' structure to deviates from an isothermal sphere. The use of a King's function assumes a spherical stellar distribution for the cluster members. It estimates the projected density of stars as a function of $R_{c}$ and $R_{t}$.

For a better spatial resolution and adequate $1\sigma$ Poisson errors the RDPs are computed by bunching the probable member stars within a set of concentric rings - the width of the rings grow with distance from the center \citep{Bica08, Camargo18, Camargo19, Camargo21, Minniti21b}.

 {\tiny
 \begin{equation} 
 \sigma(R)={\sigma}_0\left[\frac{1}{\sqrt{1+(R/R_c)^2}}-\frac{1}{\sqrt{1+(R_t/R_c)^2}}\right]^2
  \label{a}
 \end{equation}}

\section{Results and discussion}
\label{sec:3}

For illustration purpose, Fig.~\ref{z1b1} displays the spatial and kinematic distribution of stars, for a radius of $R=20'$ centered in the coordinates of NGC1605a, according to pyUPMASK.
This figure reveals that both distributions show a large spread, but some overdensities can be seen, especially in the PM space. It may indicate that there are more than one comoving groups of stars within the NGC1605 projected area.

After extracting the Gaia EDR3 data to a larger area ($R=60'$), the ages for the two OCs are derived by fitting PARSEC isochrones to the CMDs (Fig.~\ref{z1c}) built with the most probable members of NGC1605a (\textit{left panels}), NGC1605b (\textit{middle panels}), and a combination of both (\textit{right panels}). The stellar content selected by the set of tools can be fitted by two isochrones with ages of $600\pm100$ Myr (NGC1605b) and $2\pm0.2$ Gyr (NGC1605a) giving an heliocentric distance for the two OCs of $d_{\odot}=2.75\pm0.40$ kpc, which are the best solutions obtained in C21. The isochrone fitting for NGC1605a is more flexible, but the derived distance remain within the uncertainty range. 
Aiming to highlight the stellar sequences on the CMDs and emphasize the physics nature of both clusters, it is implemented a KNN smoothing algorithm (Sect. \ref{sec:2a}). The heatmaps are built smoothing on 3 (\textit{top panels}) and 8 (\textit{bottom panels}) neareast neighbors to show details on the stellar sequences and the more prominent ones, respectively. These heatmaps show that the stellar sequences of NGC1605a are prominent even in the CMDs of NGC1605b.
Thus, Fig.~\ref{z1c} reinforces the idea that there are two stellar populations in the region of NGC1605.

\begin{figure}
\centering
\begin{minipage}[b]{1.0\linewidth}

\resizebox{\hsize}{!}{\includegraphics{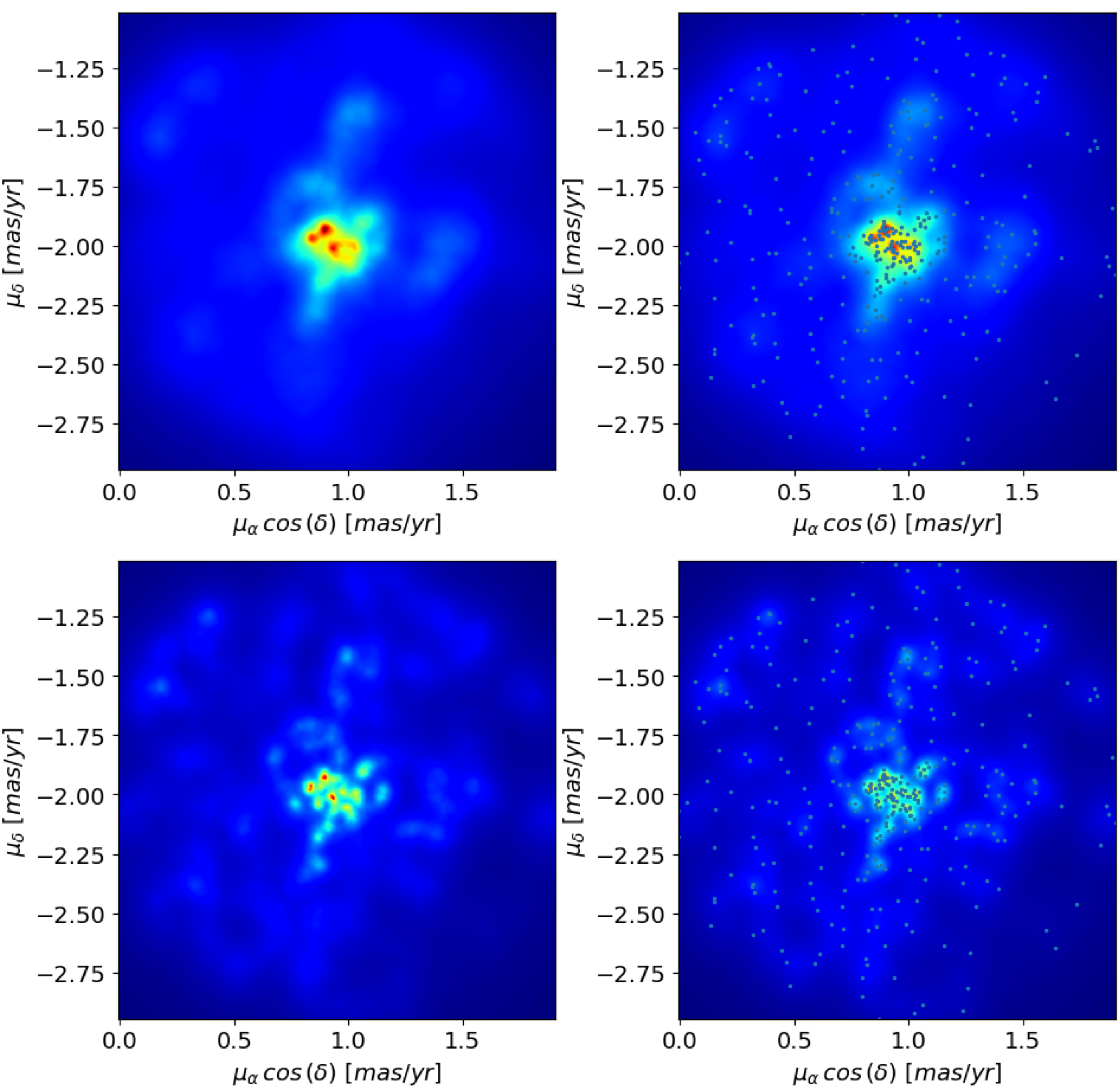}}

\caption{Gaia-EDR3 PM distribution, for NGC1605b, highlighted by the KNN-smoothing algorithm for $K=12$ (\textit{top panels}) and $K=5$ neighbors (\textit{bottom panels}). The black circles are the stars in the cluster memberlist.}
\label{g3}
\end{minipage}
\end{figure}

ASteCA provides the binarity probability for all stars on the clusters membership lists (Fig.~\ref{z1b}). The bulk of probable binary stars - even in the CMD built with Gaia-EDR3 extraction centered in NGC1605b (right panel) - are located in the region corresponding to the NGC1605a upper MS.  The upper MS is the region on the CMD of a star cluster where it is expected a large concentration of multiple systems, with the binary fraction decreasing towards the low MS \citep{Belokurov20, Penoyre22a, Penoyre22b}. 
Thus, the distribution of the probable binary stars on the CMDs (Fig.~\ref{z1b})  supports the idea that these objects compose an old binary cluster, since the bulk of binary stars are located in the region that corresponds to the upper main sequence of NGC 1605a, reinforcing its physical nature.

Figs.~\ref{z1c} and ~\ref{z1b} unveil that most stars in the memberlists of the two OCs are the same sources, which is expected since these stars differ from the Galactic field stars, which are the outliers removed by the algorithms. 
Further, as proposed by C21, NGC1605a and NGC1605b possibly make up an old binary cluster at an advanced stage of merger, so they differ significantly in ages, but practically overlap in spatial and kinematic phase space.
It turns out that the set of machine learning adopted in this work do not uses ages in the members selection and, thus, these tools are unable to efficiently isolate the stars from each cluster, since they are too close to each other in the 5D phase space \citep{Hunt23}. Since the MS of both OCs overlap in the CMDs (Fig.~\ref{z1c}), color-magnitude filters also are unable to completely isolate the clusters' sequences. On the other hand, these approaches are able to accurately isolate the binary cluster members from the field stars. For brevity, from here on in most features only the probable members for the extraction centered in the NGC1605b will be used, since it includes most of the members for the two OCs and the presence of NGC1605a on the companion's features is a robust indication in favor of its cluster nature.

The spatial distribution of stars for the old binary cluster candidate centered in the NGC1605b coordinates is shown in the Fig.~\ref{g}. 
The presence of at least two overdensities in the region is evident.

\begin{figure}
\centering
\begin{minipage}[b]{1.0\linewidth}
\begin{minipage}[b]{0.98\linewidth}
\includegraphics[width=\textwidth]{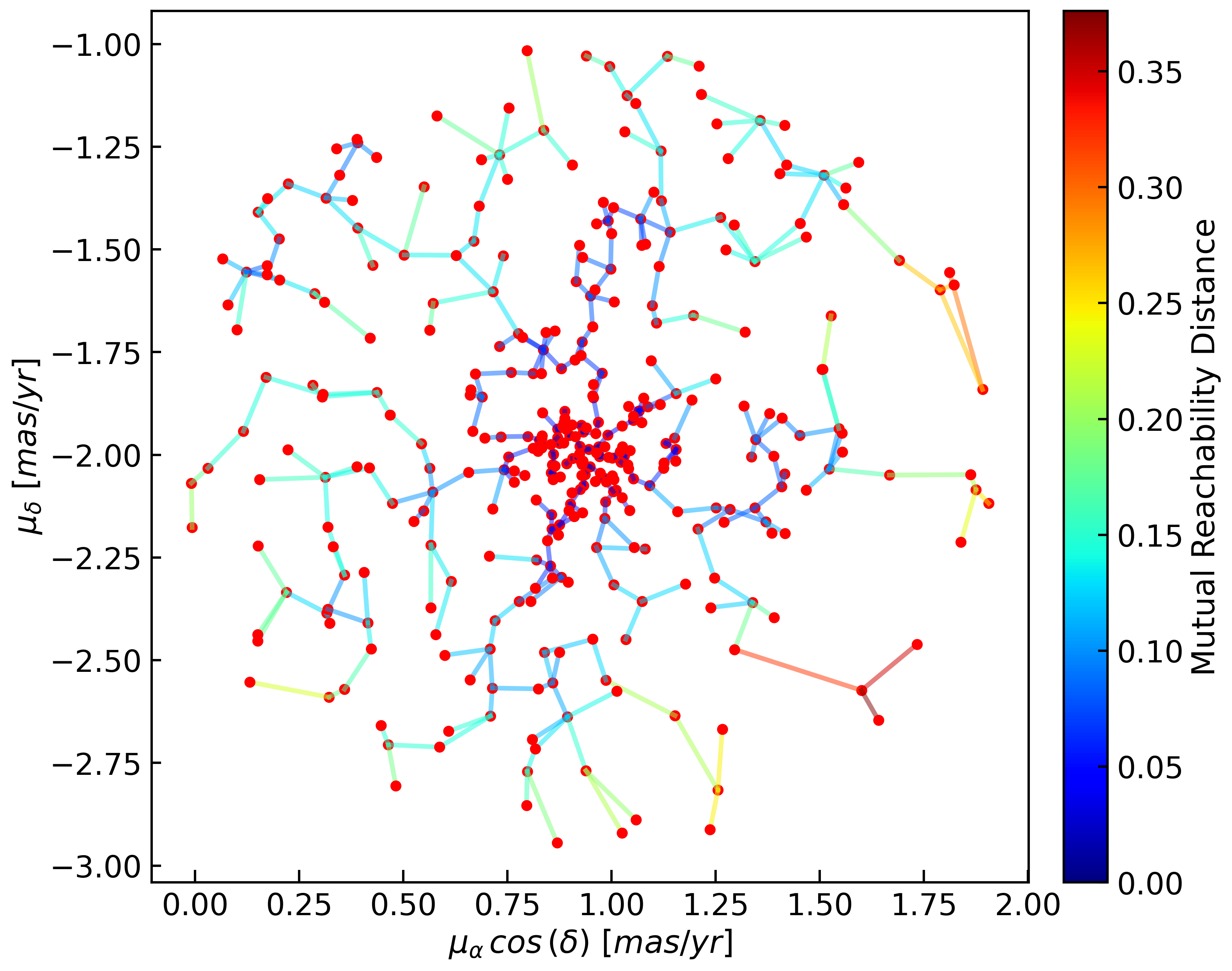}
\end{minipage}\hfill
\begin{minipage}[b]{0.98\linewidth}
\resizebox{\hsize}{!}{\includegraphics{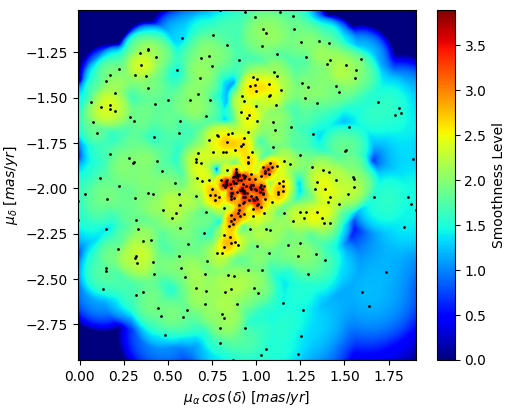}}
\end{minipage}\hfill
\caption{\textit{Top panel}: MST for the NGC1605b memberlist implemented by a combination of KNN and HDBSCAN. \textit{Bottom panel}: a zoom in the PM diagram built by using the KNN-smoothing algorithm.}
\label{g4d}
\end{minipage}
\end{figure}

In Fig.~\ref{z1y} is shown the parallax diagram for the binary cluster member-stars in the NGC1605b extraction, while the uncertainties associated to the parallaxes are displayed in Fig.~\ref{f16}.
The inverse of parallax corrected for the zero-point offset provides the cluster's distance.
The median parallax for both OCs is $\omega\,=\,0.37\pm0.04\,mas$, which corresponds to an heliocentric distance of $\sim2.7$ kpc that agree with the value obtained by isochrone fitting. 
The parallax and PMs associated uncertainties are computed as the median absolute deviation, which is less affected by missing data and outliers than other methods.

\begin{figure*}
\centering
\begin{minipage}[b]{0.7\linewidth}
\includegraphics[width=\textwidth]{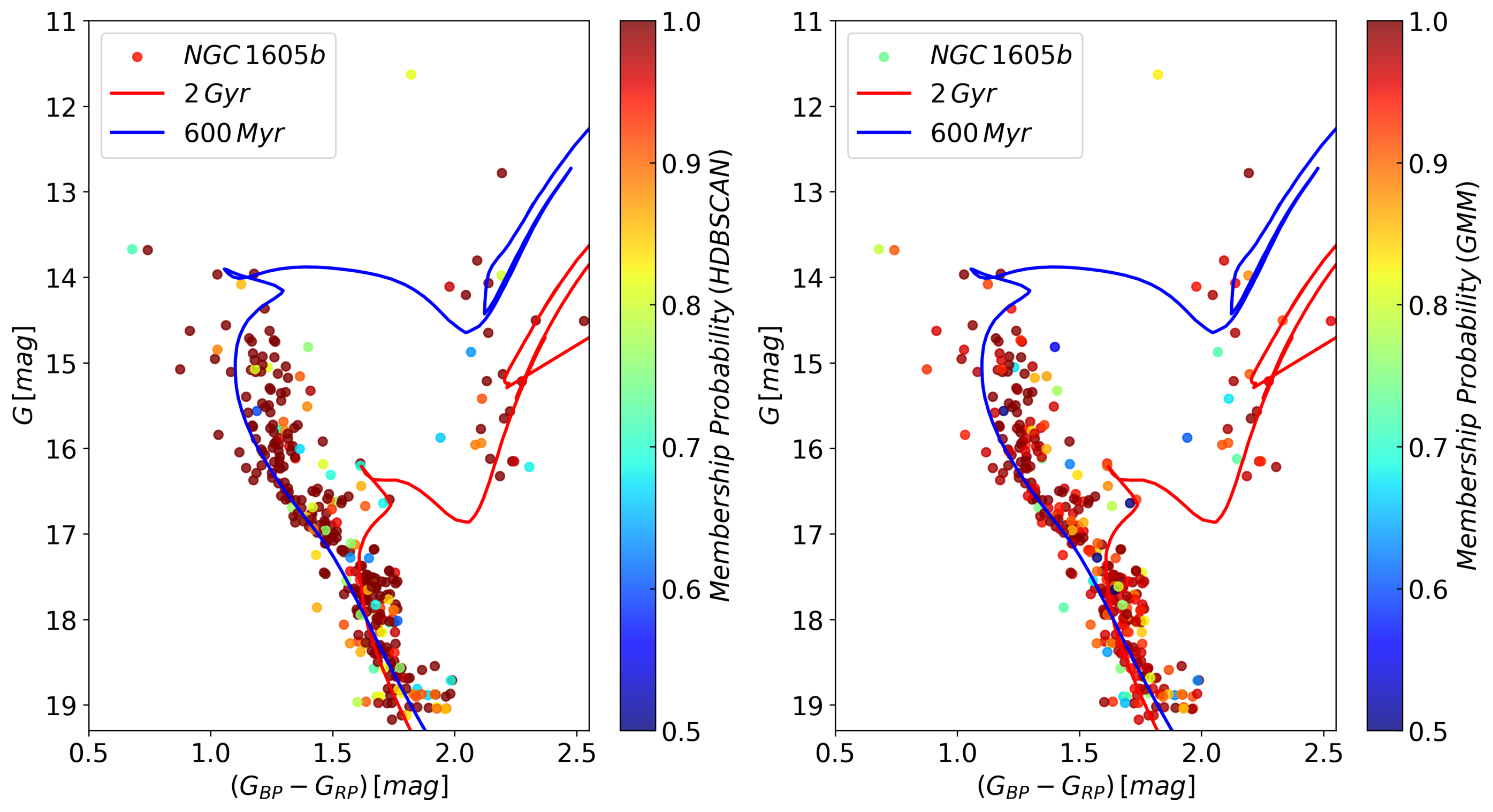}
\end{minipage}\hfill
\caption{Gaia CMDs for the NGC1605 memberlist color coded by the HDBSCAN and GMM membership probabilities.}
\label{z1a}

\end{figure*}

The median PM along RA is $0.91\pm0.05\,mas\,{yr}^{-1}$ for NGC1605a and $0.92\pm0.05\,mas\,{yr}^{-1}$ for NGC1605b while the median PM along Dec is $-1.99\pm0.04\,mas\,{yr}^{-1}$ for NGC1605a and $-1.98\pm0.04\,mas\,{yr}^{-1}$ for NGC1605b.
These values are consistent with the result of \citet{Gokmen23} for NGC1605. 
The Fig.~\ref{g3} displays the proper motion diagram for stars within the binary cluster region (NGC1605b extraction), which reinforces the presence of two major structures (top panels). This feature reveals that the stellar distribution in the region is substructured not only spatially, but kinematically as well (bottom panels). Thus, for clarity purposes the top panel of Fig.~\ref{g4d} presents the minimum spanning tree (MST) for the old binary cluster probable members. 
It uses a KNN-HDBSCAN combination along with the concepts of core distance and mutual reachability distance. For comparison, the KNN-smoothing algorithm also provides a zoom in the PM diagram (bottom panel). The heatmap is color coded by the smoothness level. Basically, the mutual reachability distance and the smoothness level indicate the degree of similarity between the stars.

\begin{figure*}
  \centering
  \begin{minipage}[b]{0.7\linewidth}
    \includegraphics[width=0.98\linewidth]{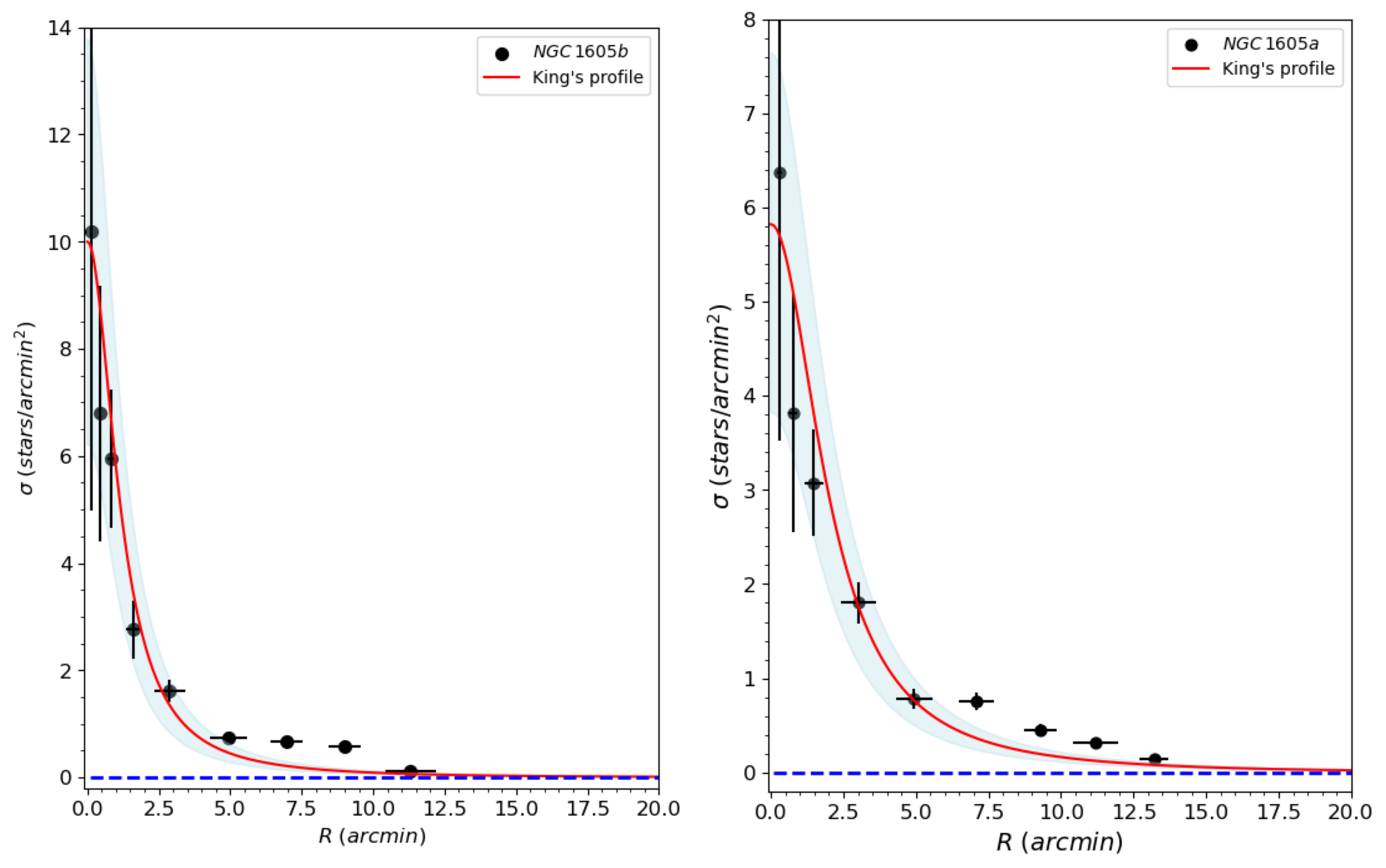}%
  \end{minipage}\hfill
  \caption{RDP for NGC1605b and NGC1605a. Blue region: $1\sigma$ King fit uncertainty.}
    \label{rdp1}
\end{figure*}

\begin{figure*}
\centering
\begin{minipage}[b]{0.90\linewidth}
\begin{minipage}[b]{0.48\linewidth}
\resizebox{\hsize}{!}{\includegraphics{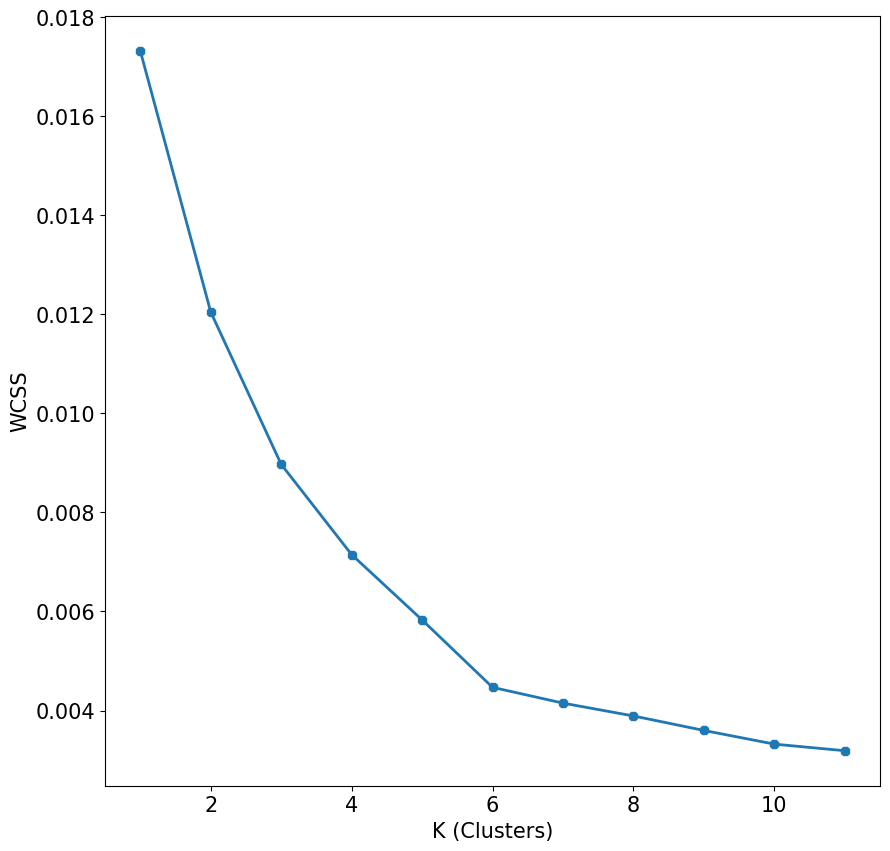}}
\end{minipage}\hfill
\begin{minipage}[b]{0.48\linewidth}
\includegraphics[width=\textwidth]{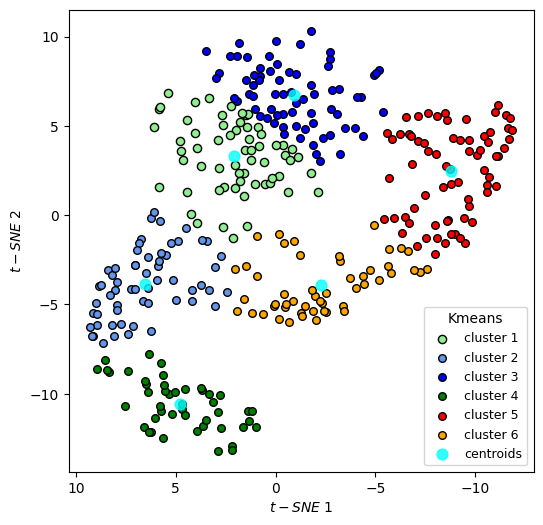}
\end{minipage}\hfill
\begin{minipage}[b]{0.48\linewidth}
\resizebox{\hsize}{!}{\includegraphics{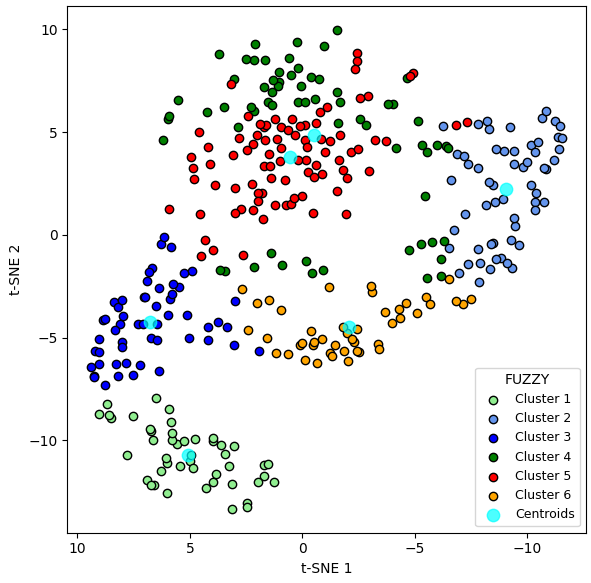}}
\end{minipage}\hfill
\begin{minipage}[b]{0.48\linewidth}
\includegraphics[width=\textwidth]{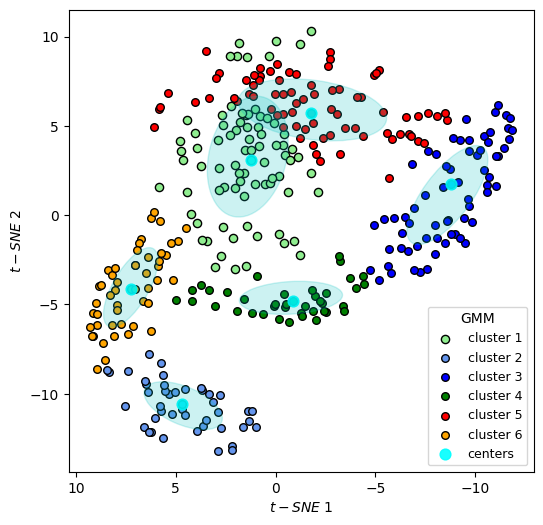}
\end{minipage}\hfill
\end{minipage}
\caption{In the \textit{Top-Left panel} is shown the result of applying the PCA elbow method, which found 6 clusters in the NGC 1605 area. The wcss (within-clusters sum-of-squares) measures the sum of squared distances between each star and its assigned cluster center as a function of the number of clusters. The t-SNE 2D projection (perplexity=50) of the stars in the NGC 1605b memberlist clustered by Kmeans (\textit{Top-right panel}), Fuzzy (\textit{Bottom-Left panel}), and GMM (\textit{Bottom-right panel}).
}
\label{z1x}

\end{figure*}

Membership probabilities for the OCs probable members are assigned by using HDBSCAN, GMM, ASteCA, and pyUPMASK. For brevity, the Fig.~\ref{z1a} shows the CMDs for NGC1605b color coded by the HDBSCAN and GMM membership probabilities. Once more, the presence in the region of an old stellar population that characterizes NGC1605a is evident. As an age effect, NGC1605a has a more prominent RGB than its counterpart NGC1605b.

It is important to highlight that most NGC1605a main sequence stars are fainter than the magnitude limit ($G=18\,mag$) that is commonly applied as quality cut in the Gaia's astrometry. The application of strict quality filters may make some old and/or distant clusters like NGC1605a undetectable, as these tools remove most of their stars \citep{Piatti22, Hunt23}. The parallax quality cut also affect these kind of objects, particularly for OCs located more than a few kpc away \citep{Chulkov22, Majaess25}.

The RDPs for NGC1605a and NGC1605b (Fig.~\ref{rdp1}) present a conspicuous excess over the king-like profile in the OCs outskirts. Such a feature also was detected in C21 and seems to be the contribution of the other pair member cluster. However, the best-fitting solutions superimposed on the OCs RDPs show that the clusters members are spatially distributed over a wide area. NGC1605a presents a $R_t\approx47.5\,arcmin$ while for NGC1605b the value is  $R_t\approx42.7\,arcmin$. The core radius of NGC1605a is $R_c=2.1\pm0.2\,arcmin$ and the central density ${\sigma}_0=5.8\pm0.2\,stars/arcmin$. The structural parameters for NGC1605b are $R_c=1.2\pm0.2\,arcmin$ and ${\sigma}_0=10.1\pm0.6\,stars/arcmin$. ASteCA individually provides similar values for the clusters' structural parameters.

In turn, to help visualize the stellar distribution in the high-dimensional phase space, a t-SNE algorithm is used to convert the previously normalized Gaia-EDR3 data from a 5D space to a 2D map. Then, the Kmeans (both, classical and Fuzzy implementations) and GMM are employed to group the data into 6 clusters as suggests the PCA elbow method (Fig.~\ref{z1x}). 
Interestingly, the elbow method provides the same number of clusters found in C21 that suggests an ongoing merger event between two clusters during a close encounter, and four subclusters as possible products of the merger process.
Although, care is recommended in the interpretation of t-SNE graphs, these features support the results obtained by C21.

Such a scenario could be produced by a close encounter between a cluster and a cluster group formed from the same giant molecular cloud. However, as pointed out by C21, it is extremely unlikely that such a system has survived for so long. C21 raises two possible interpretations for this scenario  \textit{i)} tidal ejected stars may coalesce together forming epicyclic overdensities along the tidal tails \citep{Jerabkova21}, or \textit{ii)} both clusters are being fragmented into subclusters by drag forces during the merging process.

If there is a final merger product of two clusters after a close encounter, it may be a single cluster coherent spatially and kinematically, but with two different ages. 
It seems to be the case for NGC1605a and NGC1605b. The distribution of the stellar content of these OCs in the 5D phase space points to an old binary cluster at an advanced stage of merger.

\section{Concluding remarks}
\label{sec:4}

This work made use of a set of well established machine learning algorithms (pyUPMASK, ASteCA, kmeans, GMM, and HDBSCAN) applyed to the Gaia-EDR3 5D astrometric space to analyse the nature of the old binary cluster candidate NGC1605a and NGC1605b. The combination of these algorithms leads to more robust results, since the limitations of one of them may be covered by the others.
In addition, a KNN smoothing algorithm is used to enhances the visualization of features like stellar overdensities in the multi-dimensional Gaia data, as well as, the clusters' intrinsic stellar sequences on the CMDs.

The clusters' fundamental parameters derived are an age of $2\pm0.2$ Gyr for NGC1605b and $600\pm100$ Myr for NGC1605a. Both are located at a heliocentric distance of $\sim2.7$ kpc in accordance with previous results (C21). 
Structural parameters show that these OCs extend along a wide overlapping area with tidal radii larger than $40\, arcmin$.

The age, size, kinematic, and location of NGC1605a and NGC1605b are consistent with an ongoing merger scenario after a tidal capture during a close encounter of two clusters at different evolutionary stages, as argues \citet{Camargo21}.  
In this sense, the Gaia data was filtered by selecting sources within a proper motion radius of $1\,mas\,{yr}^{-1}$ of the clusters' centers, which is a stric quality cut even for a single cluster.
Even so, it is possible to notice the presence of more than one object in all the features shown in this study.

The PCA elbow method found 6 clusters in the binary cluster candidate area and, thus, the t-SNE algorithm reduces the normalized Gaia-EDR3 data from a 5D space to a 2D map and combined with Kmeans, Fuzzy, and GMM identify the 6 clusters.

It seems that the merger end-product of NGC1605a and NGC1605b will be a single OC (NGC1605) coherent spatially and kinematically, but with two different ages.

The MMLM reinforces C21 results and, in addition, proven to be highly effective in uncovering hidden clusters and their substructures, which highlights its potential for future applications, particularly in the analysis of complex stellar structures.

\vspace{0.8cm}

\textit{Acknowledgements}: 
The author thank the anonymous referee for constructive comments and suggestions that greatly strengthened the manuscript. 

This work has made use of data from the European Space Agency (ESA)
mission {\it Gaia} (\url{https://www.cosmos.esa.int/gaia}), processed by
the {\it Gaia} Data Processing and Analysis Consortium (DPAC,
\url{https://www.cosmos.esa.int/web/gaia/dpac/consortium}). Funding
for the DPAC has been provided by national institutions, in particular
the institutions participating in the {\it Gaia} Multilateral Agreement.



\end{document}